\documentclass[10pt,a4paper]{article}
\usepackage[top=2.5cm,left=2.5cm,right=2.5cm, footskip=1.5cm]{geometry}

\usepackage{amsmath,amssymb}
\usepackage{siunitx}

\usepackage{changepage}

\usepackage[utf8]{inputenc}

\usepackage{textcomp,marvosym}

\usepackage[table]{xcolor}
\usepackage{float}

\usepackage{nameref}

\usepackage[colorlinks=true]{hyperref}
\definecolor{hrcolor}{rgb}{0.5,0.1,0.3}
\hypersetup{linkcolor=hrcolor}
\hypersetup{citecolor=hrcolor}
\hypersetup{urlcolor=hrcolor}
\usepackage[right]{lineno}

\usepackage{microtype}
\DisableLigatures[f]{encoding = *, family = * }

\usepackage[aboveskip=1pt,labelfont=bf,labelsep=period,singlelinecheck=off]{caption}

\usepackage[
style=numeric-comp,
sortcites,
sorting=none,
url=false,
isbn=false,
doi=false,
maxnames=2,
eprint=false]{biblatex}
\bibliography{biblio}
\AtEveryBibitem{
 \clearlist{language}
 \clearlist{editor}
 \clearfield{note}
 \clearfield{day}
 \clearfield{month}
 \clearfield{pages}
}

\DeclareCiteCommand{\textcite}
  {\usebibmacro{prenote}}
  {\usebibmacro{citeindex}%
   \printnames{labelname}%
   \setunit{\addspace}%
   \printtext[parens]{\printfield{year}}
   \setunit{\addspace}%
   \printtext[brackets]{\bibhyperref{\printfield{labelnumber}}}}
  {\multicitedelim}
  {\usebibmacro{postnote}}

\usepackage{graphicx}

\usepackage{multicol}

\usepackage{calc}

\usepackage{graphicx}
\usepackage{tcolorbox}
\tcbuselibrary{skins, theorems}
\usepackage{nameref}            

\definecolor{boxTitle}{rgb}{0.80, 0.80, 0.9}       
\definecolor{boxBackground}{rgb}{0.98, 0.98, 0.99}
\definecolor{boxFrame}{rgb}{0.80, 0.80, 0.90}      

\tcbset{my box/.style={
    enhanced, fonttitle=\bfseries,
    colback=boxBackground, colframe=boxFrame,
    coltitle=black, colbacktitle=boxTitle,
    attach boxed title to top left={xshift=0.3cm,
                                    yshift*=-\tcboxedtitleheight/2},
    boxed title style={
    }
  }
}

\newtcolorbox[auto counter]{infobox}[2][]{%
  my box, title={Box~\thetcbcounter: #2}, #1}

\usepackage{csquotes}

\newcommand{\calcium}{Ca$^{2+}$}

\title{Representation learning in cerebellum-like structures}

\author{Lucas Rudelt\textsuperscript{1}\thanks{\href{mailto:lucas.rudelt@ds.mpg.de}{lucas.rudelt@ds.mpg.de}} , Fabian A Mikulasch\textsuperscript{1,2}\thanks{\href{mailto:fabian.mikulasch@ds.mpg.de}{fabian.mikulasch@ds.mpg.de}} , Viola Priesemann\textsuperscript{1,3,4}\thanks{\href{mailto:viola.priesemann@ds.mpg.de}{viola.priesemann@ds.mpg.de}} , André Ferreira Castro\textsuperscript{5}\thanks{\href{mailto:andre.ferreira-castro@tum.de}{andre.ferreira-castro@tum.de}} \\
\small
\textbf{1} Max-Planck-Institute for Dynamics and Self-Organization, Göttingen, Germany
\\
\small
\textbf{2} Friedrich Miescher Institute for Biomedical Research, Basel, Switzerland
\\
\small
\textbf{3} Campus Institute for Dynamics of Biological Networks, Georg-August University, Göttingen, Germany
\\
\small
\textbf{4} Bernstein Center for Computational Neuroscience (BCCN), Göttingen, Germany
\\
\small
\textbf{5}  School of Life Sciences, Technical University of Munich, Freising, Germany
\\
\small}

\date{}
 
\begin{document}
\maketitle

\begin{abstract}
Animals use past experiences to adapt future behavior. To enable this rapid learning, vertebrates and invertebrates have evolved analogous neural structures like the vertebrate cerebellum or insect mushroom body. 
A defining feature of these circuits is a large expansion layer, which re-codes sensory inputs to improve pattern separation, a prerequisite to learn non-overlapping associations between relevant sensorimotor inputs and adaptive changes in behavior.
However, classical models of associative learning treat expansion layers as static, assuming that associations are learned through plasticity at the output synapses. 
Here, we review emerging evidence that also highlights the importance of plasticity within the expansion layer for associative learning. 
Because the underlying plasticity mechanisms and principles of this representation learning are only emerging, we systematically compare experimental data from two well-studied circuits for expansion coding---the cerebellum granule layer and the mushroom body calyx.
The data indicate remarkably similar interneuron circuits, dendritic morphology and plasticity mechanisms between both systems that hint at more general principles for representation learning. 
Moreover, the data show strong overlap with recent theoretical advances that consider interneuron circuits and dendritic computations for representation learning. 
However, they also hint at an interesting interaction of stimulus-induced, non-associative and reinforced, associative mechanisms of plasticity that is not well understood in current theories of representation learning. 
Therefore, studying expansion layer plasticity will be important to elucidate the mechanisms and full potential of representation learning for behavioral adaptation.
\end{abstract}

\section*{A circuit for rapid learning of adaptive behavior}

A central role of the brain is to adapt behavior based on relevant sensorimotor information. Some predictive cues, such as circadian rhythms, sexual pheromones, or ecological signals, are stable enough across evolution so that their processing can be genetically hardwired~\cite{allada1998mutant, blankers2021sex, bargiello1984restoration}. However, for exploratory species and complex behaviors, the relevance of sensory information is dynamic and context-dependent, requiring flexible, experience-dependent learning mechanisms~\cite{heald2023computational, kumano2016context, birman2019flexible,sousa2025multidimensional}. Across diverse taxa, animals need to associate sensory stimuli with biologically meaningful outcomes—such as an electric shock or a motor error signal—often after a single trial~\cite{hammer1993identified, hammer1995learning, adam_fast_2022}. Understanding how neural circuits support this rapid associative learning remains a central question in neuroscience.

\begin{figure}
\centering
\includegraphics[scale=0.8]{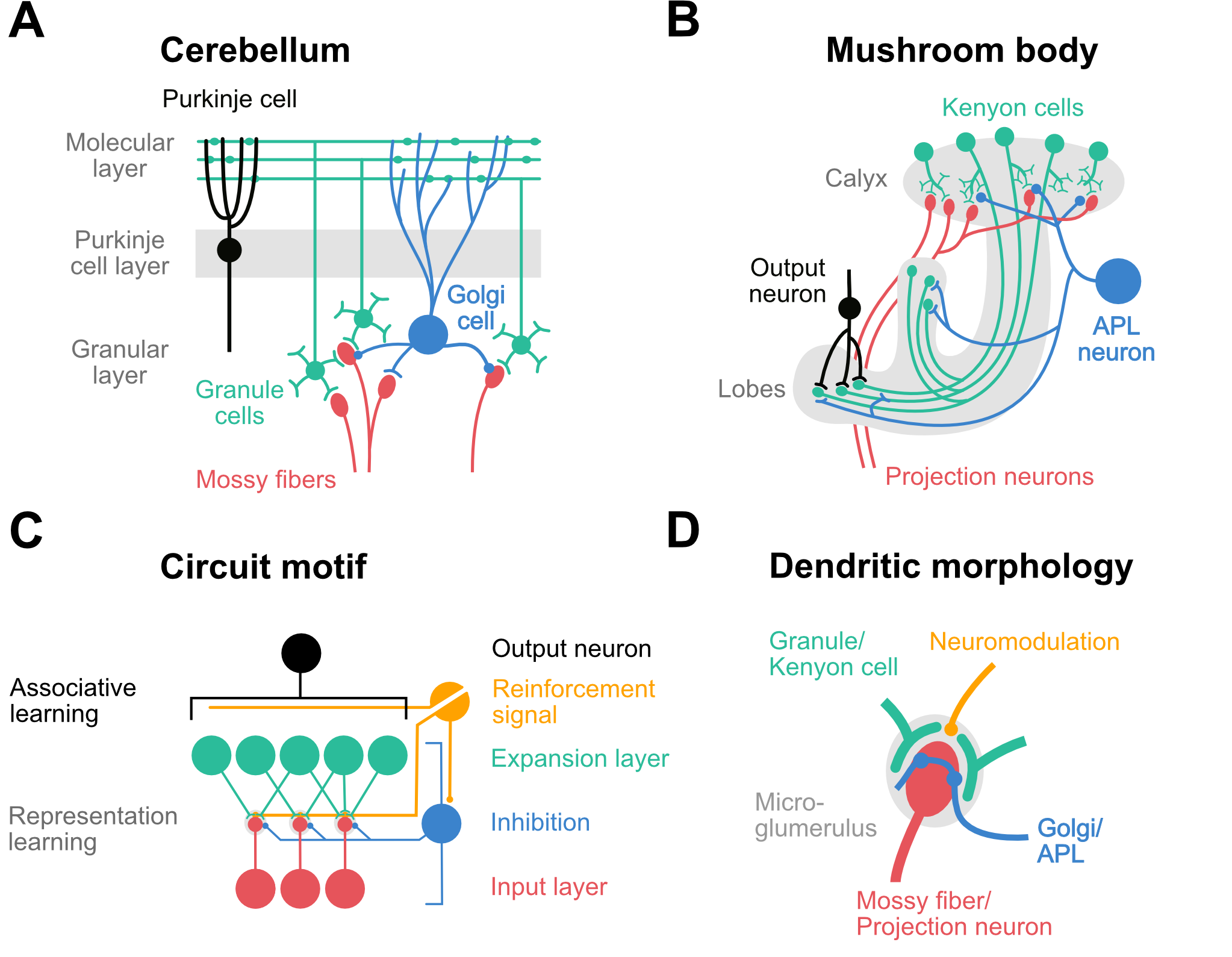} 
\caption{\textbf{Analogous circuit motifs and dendritic morphology in cerebellum-like structures.} 
\textbf{(A)} Cerebellum circuit as shown in~\cite{cayco2019re}. In the cerebellum, mossy fibers convey inputs to granule cells in the granular layer. These cells generate a high-dimensional input representation, shaped by feedforward and feedback inhibition by Golgi cells. Outputs from the granular layer converge onto Purkinje cells that modulate behavior and adapt motor control.
\textbf{(B)} In the mushroom body, these functions are implemented by projection neurons (input layer), Kenyon cells in the calyx (expansion layer), the APL in Drosophila (inhibitory interneuron), and mushroom body output neurons in the lobes. Circuit shown as in~\cite{cayco2019re}.
\textbf{(C)} 
Both systems thus implement an analogous circuit motif where inputs are re-coded in a divergent expansion layer of a much higher number of principal neurons, whose outputs again converge onto a much smaller number of output neurons. Here, climbing fibers in the cerebellum and neuromodulatory inputs in the mushroom body convey reinforcement signals to drive associative learning. 
In addition, climbing fibers and neuromodulatory inputs can directly or indirectly affect responses and plasticity of principal or inhibitory neurons in the expansion layer, thus potentially enabling representation learning \cite{leiss2009synaptic, boto2019independent, qiao_input-timing-dependent_2022, cao2022nicotine}. 
\textbf{(D)} Input synapses and principal dendrites in both systems form microglomeruli, where dendritic elaborations from many principal cells enwrap a single synaptic bouton. Microglomeruli are additionally innervated by inhibitory axons as well as neuromodulatory inputs, thus providing a structural unit for efficient synaptic transmission, localized computation and plasticity (main text).}
\label{fig:curcuit_overview}
\end{figure}

Despite large differences in brain architecture, many distantly related taxa have converged on a common circuit motif for rapid associative learning. These so-called \textit{cerebellum-like structures}---including the cerebellum itself and the insect mushroom body (\textbf{MB}), among others---transform sensory inputs through a feedforward expansion layer, where relatively few input fibers are recoded into large populations of principal neurons, whose axons again converge onto a much smaller population of output neurons~\cite{eccles1967cerebellum, cayco2019re, strausfeld2002organization, aso2014neuronal, eichler2017complete,dorkenwald2024neuronal, nguyen2023structured} (Figure~\ref{fig:curcuit_overview}). Classical theories, from Marr and Albus~\cite{rosenblatt1958perceptron,marr1969theory, albus1971theory} to adaptive filtering frameworks~\cite{ito1970neurophysiological, fujita1982adaptive, dean2010cerebellar, sejnowski1977storing}, emphasized that such expansion enhances pattern separation~\cite{eccles1967cerebellum, cayco2019re, strausfeld2002organization, aso2014neuronal, litwin2017optimal, cayco2017sparse, cayco2019re} and supports associative learning by enabling downstream synapses to link stimuli with behavioral relevance signals~\cite{eccles1967cerebellum, owald2015activity, bouzaiane2015two, arican2023mushroom}. In this view, expansion layers were treated as static recoding devices, while plasticity was confined to output synapses where climbing fibers or dopaminergic neurons guide learning~\cite{ito1982climbing, ito1982long, kim2007d1, qin2012gamma, cohn_coordinated_2015}.

Recent work has gone further by demonstrating that the expansion layers themselves undergo experience-dependent plasticity, with expansion layer responses adapting across multiple timescales~\cite{hansel2001beyond, d2009timing, kremer2010structural, gao2012distributed, d2016distributed, pali2024understanding, Baltruschat2020}. These findings point to a new perspective in which expansion layers are not fixed encoders but dynamically adapt to their inputs. What remains unclear, however, are the functional principles by which such plasticity shapes input representations to support associative learning and adaptive behavior.

To address these questions, this review focuses on the cerebellum and the insect MB—two systems that provide uniquely tractable models for dissecting the mechanisms of representations learning in expansion layers. Both systems separate input representations from output stages, enabling precise analysis of how representations are formed and used downstream to adapt behavior \cite{eichler2017complete, nguyen2023structured}. Furthermore, their structural simplicity, well-characterized computational function, and accessibility to both genetic and physiological interrogation make them ideal substrates for uncovering principles of learning across brains. We highlight how cellular features such as dendritic integration and inhibitory feedback provide the substrate for efficient learning, synthesize evidence for plasticity within these layers, and relate it to theories of representation learning and predictive coding. 
Moreover, by relating plasticity to behavioral experiments and modeling studies, we identify common functional principles of representation learning in expansion layers.
Finally, we conclude with open questions and experimental predictions that connect cellular mechanisms to learning theory.

\section*{Analogous architecture and synapto-dendritic organization}

To support adaptive learning and flexible behavior, both the cerebellum and the mushroom body (MB) implement an analogous circuit motif: an expansion coding architecture that transforms lower-dimensional input into sparse, high-dimensional representations~\cite{albus1971theory,marr1969theory,ito1964cerebellar,litwin2017optimal}. In both systems, this is achieved by a massive increase in the number of principal neurons at the input stage—granule cells (\textbf{GrCs}) in the cerebellum and Kenyon cells (\textbf{KCs}) in the MB—each receiving input from a small subset of afferents, mossy fibers (\textbf{MFs}) and projection neurons (\textbf{PNs}), respectively (Figure~\ref{fig:curcuit_overview}A,B). Input connectivity in both systems is sparse and largely unstructured, with GrCs receiving input from on average $\sim 4$ MFs and KCs from $\sim 7$ PN boutons (in the adult fly)~\cite{dorkenwald2024neuronal, eichler2017complete, nguyen2023structured, schlegel2024whole}. Combined with threshold nonlinearities in GrCs and KCs, this supports a high-dimensional coding space \cite{litwin2017optimal}. These representational features have been shown to facilitate pattern separation, which is essential for forming distinct associations at the convergent output stage, Purkinje cells in the cerebellum and output neurons (\textbf{MBONs}) in the MB~\cite{campbell_imaging_2013, lin2014sparse, cayco2017sparse, jeanne_convergence_2015, endo_synthesis_2020, shadmehr_population_2020}. 

Inhibitory interneurons further shape the input representation. In the cerebellum, Golgi cells (\textbf{GoCs}) provide both feedforward and feedback inhibition to GrCs, while in the \emph{Drosophila} MB, the anterior paired lateral (\textbf{APL}) neuron serves a similar function for KCs (Figure~\ref{fig:curcuit_overview}A,B). These inhibitory motifs ensure that principal cell responses are sparse and decorrelated~\cite{laurent2002olfactory, mitchell2003shunting, mapelli2010high, papadopoulou2011normalization, lin2014sparse, stopfer2014central, prisco2021anterior, fleming2024}, they normalize responses~\cite{prisco2021anterior}, and control temporal response dynamics~\cite{gandolfi2013theta, kanichay2008synaptic, mapelli2007spatial, d2008critical, d2009timing, mapelli2010combinatorial}. This combination of expansion and inhibition—illustrated abstractly in Figure~\ref{fig:curcuit_overview}C—is a hallmark of cerebellum-like architectures that has emerged across phyla~\cite{Modi2020}.

Beyond their shared circuit logic, both the cerebellar granule cells and MB Kenyon cells possess strikingly similar dendritic morphologies. Both cell types form microglomerular structures, where clawed dendritic specializations enwrap the terminals of mossy fiber or projection neuron afferents within glial-lined microglomeruli, in which excitatory, inhibitory, and neuromodulatory inputs converge onto dendritic specializations of GrCs or KCs (Figure~\ref{fig:curcuit_overview}D)~\cite{leitch1996gabaergic,strausfeld2002organization, yasuyama2002synaptic, frambach2004f,leiss2009synaptic, kanichay2008synaptic, sinakevitch2013apis, prestori2013gating}. These microdomains allow for efficient synaptic transmission shaped by inhibitory or neuromodulatory inputs~\cite{digregorio2002spillover, kanichay2008synaptic, leiss2009synaptic, prestori2013gating, qiao_input-timing-dependent_2022}. 
Moreover, the convergence of all these learning-relevant inputs at a synapse, as well as high concentrations of post-synaptic f-actin at the dendrites~\parencite{capani2001filamentous, frambach2004f,leiss2009synaptic}, make the microglomeruli ideal substrates for plasticity to support representation learning.

\section*{Plasticity in expansion layers}

The striking similarity in microglomerular structure between the cerebellum and MB motivates a closer examination of the synaptic, inhibitory, and neuromodulatory mechanisms that support plasticity in the expansion layer.
In the following sections, we outline how stimulation-induced and reinforcement-dependent mechanisms modify excitatory and inhibitory connectivity in the cerebellar granule layer and mushroom body calyx. 
Together, these forms of plasticity reveal how expansion circuits flexibly adjust input strength to encode the salience, timing, and behavioral relevance of sensory signals.

\subsection*{Stimulation-induced plasticity of afferent inputs}

Stimulation-induced plasticity strengthens or weakens afferent synapses onto principal neurons following specific activation patterns. Long-term potentiation (\textbf{LTP}), by enhancing the efficacy of afferent inputs, can amplify salient activity patterns through increased response magnitude, reduced latency, and improved reliability~\cite{d2009timing}. In contrast, long-term depression (\textbf{LTD}) reduces input efficacy for specific inputs of principal cells. In the cerebellum, extensive work has elucidated the conditions and mechanisms underlying LTP and LTD (see \cite{pali2024understanding} for a comprehensive review), whereas in the mushroom body (MB) calyx, stimulation-induced plasticity has only recently begun to be explored (see Table~\ref{tab:stim_potentiation} for an overview of relevant studies).



In the cerebellum, LTP and LTD at mossy fiber–granule cell (\textbf{MF–GrC}) synapses require the coincident occurrence of presynaptic activity and postsynaptic depolarization that together elevate intracellular \calcium{} levels in granule cells~\cite{armano2000long, gall2005intracellular}. 
Here, LTP and LTD follow a rule similar to Bienenstock-Cooper-Munro theory~\cite{bienenstock1982theory}, in which low levels of postsynaptic \calcium{} lead to no plasticity, whereas intermediate levels lead to LTD and high levels lead to LTP~\cite{gall2005intracellular,d2009differential}. 
In cerebellar slice experiments, the conditions for LTP are achieved through several mechanisms, including
strong presynaptic firing through theta burst stimulation that is combined with postsynaptic depolarization~\cite{d1999evidence, armano2000long, maffei2002presynaptic, nieus2006ltp, d2009differential}, or repeated, precisely timed pre- and postsynaptic firing in the theta range~\cite{sgritta2017hebbian}. 
In contrast, LTD occurs for weaker MF stimulation (i.e., short high-frequency stimulation~\cite{gall2005intracellular} or tetanic low-frequency stimulation~\cite{d2009differential}), 
as well as an action potential that repeatedly precedes MF stimulation during spike-timing-dependent-plasticity (\textbf{STDP})~\cite{sgritta2017hebbian}.
Importantly, these protocols are consistent with observed \textit{in vivo} activity patterns of mossy fibers~\cite{kase1980discharges,dugue2009electrical}.
Moreover, \emph{in vivo} experiments in anesthetized rats and mice could reproduce LTP and LTD through whisker stimulation~\cite{roggeri2008tactile, lu2022facial}, suggesting that LTP and LTD play a role in the cerebellum expansion layer in relevant conditions \emph{in vivo}.

Since LTD and LTP depend on the postsynaptic depolarization, this suggests that synaptic plasticity in principal cells is highly sensitive to dendritic inhibition. Consistent with this, \emph{in vivo} facial stimulation in mice and rats induces LTP of MF-GrC synapses under pharmaceutical blocking of GABA$_\text{A}$ receptors, whereas LTD prevails when inhibition is unblocked~\cite{roggeri2008tactile, lu2022facial}. Similarly, in cerebellar slices, it has been found that LTP is confined to a small region where excitation exceeds inhibition, whereas LTD occurs in the surround where inhibition dominates~\cite{mapelli2007spatial, casali2020cellular}. Finally, unblocking of inhibition has been found to reverse the timing dependence of STDP at the MF-GrC synapse~\cite{sgritta2017hebbian}. These findings highlight the crucial role of inhibition in gating the direction of plasticity at MF–GrC synapses, thereby shaping plasticity in the cerebellar expansion layer.

Beyond a functional characterization, extensive work has delineated the molecular pathways underlying the induction of LTP and LTD at MF–GrC synapses. 
In line with the \calcium{} dependence of LTP and LTD, their induction depends on the co-activation of postsynaptic metabotropic glutamate receptors (\textbf{mGluRs}), N-methyl-D-aspartate receptors (\textbf{NMDARs}), voltage-gated calcium channels (\textbf{VGCCs}) and the mobilization of internal \calcium{} stores~\cite{rossi1996differential, d1999evidence, maffei2002presynaptic, sgritta2017hebbian}. 
Of those, NMDARs have been found to be necessary for LTP induction, highlighting its sensitivity to coincident presynaptic input and postsynaptic depolarization, while mGluRs have been found to be sufficient for LTD during STDP~\cite{sgritta2017hebbian}.
Finally, an important property of MF-GrC plasticity is that both, LTP and LTD induction, depend on nitric oxide (\textbf{NO}).
This is interesting, because NO acts a retrograde messenger to promote \emph{presynaptic} modifications~\cite{d2005long, lu2022facial}.

In line with this, LTP and LTD expression is predominantly presynaptic. During LTP, this is reflected in a reduced excitatory postsynaptic potential (\textbf{EPSP}) variability and decreased paired-pulse ratio~\cite{sola2004increased, nieus2006ltp}, which suggest an increase in presynaptic release probability. 
In contrast, LTD reverses these effects through a decrease in release probability, thus enabling the synapse to adjust its strength in either direction~\cite{sola2004increased, nieus2006ltp}. 
In addition, complementary postsynaptic contributions---such as enhanced NMDA receptor currents and increased excitability---have been observed for long-term plasticity~\cite{armano2000long, sgritta2017hebbian}. Computational models suggest that presynaptic potentiation primarily shortens response latency and enhances reliability, whereas postsynaptic potentiation boosts spike output gain, in line with experimental observations~\cite{nieus2006ltp}. 
Together, these mechanisms thus enable LTP and LTD to flexibly adjust multiple aspects of granule cell encoding.

In contrast to the cerebellar granule layer, evidence for stimulation-induced plasticity in the mushroom body (MB) calyx is only beginning to emerge.
However, a comparable bidirectional rule also seems to operate at the input synapses in the MB calyx. 
A recent study in \textit{Drosophila} reported that repeated pairings of precisely timed input and postsynaptic activation lead to pronounced LTP and LTD of PN-KC synapses at $\gamma$ Kenyon cells
~\cite{qiao_input-timing-dependent_2022}. 
In particular, PN-KC synapses undergo LTP when postsynaptic depolarization follows shortly after PN stimulation, whereas they undergo LTD when the postsynaptic depolarization precedes PN input. 
Although the detailed induction and expression mechanisms remain to be fully characterized, this plasticity depends on the dopaminergic receptor Dop2R and postsynaptic \calcium{} signaling, as blocking VGCCs abolishes the effect~\cite{qiao_input-timing-dependent_2022}. 
Thus, similar calcium-dependent coincidence mechanisms may operate at both cerebellar and MB input stages.
However, the induction window in the MB (at least \qty{-40}{\milli\second} to \qty{100}{\milli\second})~\cite{qiao_input-timing-dependent_2022} is substantially broader than in the cerebellum (about -\qty{25}{\milli\second} to \qty{25}{\milli\second})~\cite{sgritta2017hebbian}, which may reflect the slower timescales of olfactory processing in the MB calyx. 

\begin{figure}
\centering
\includegraphics[width=\textwidth]{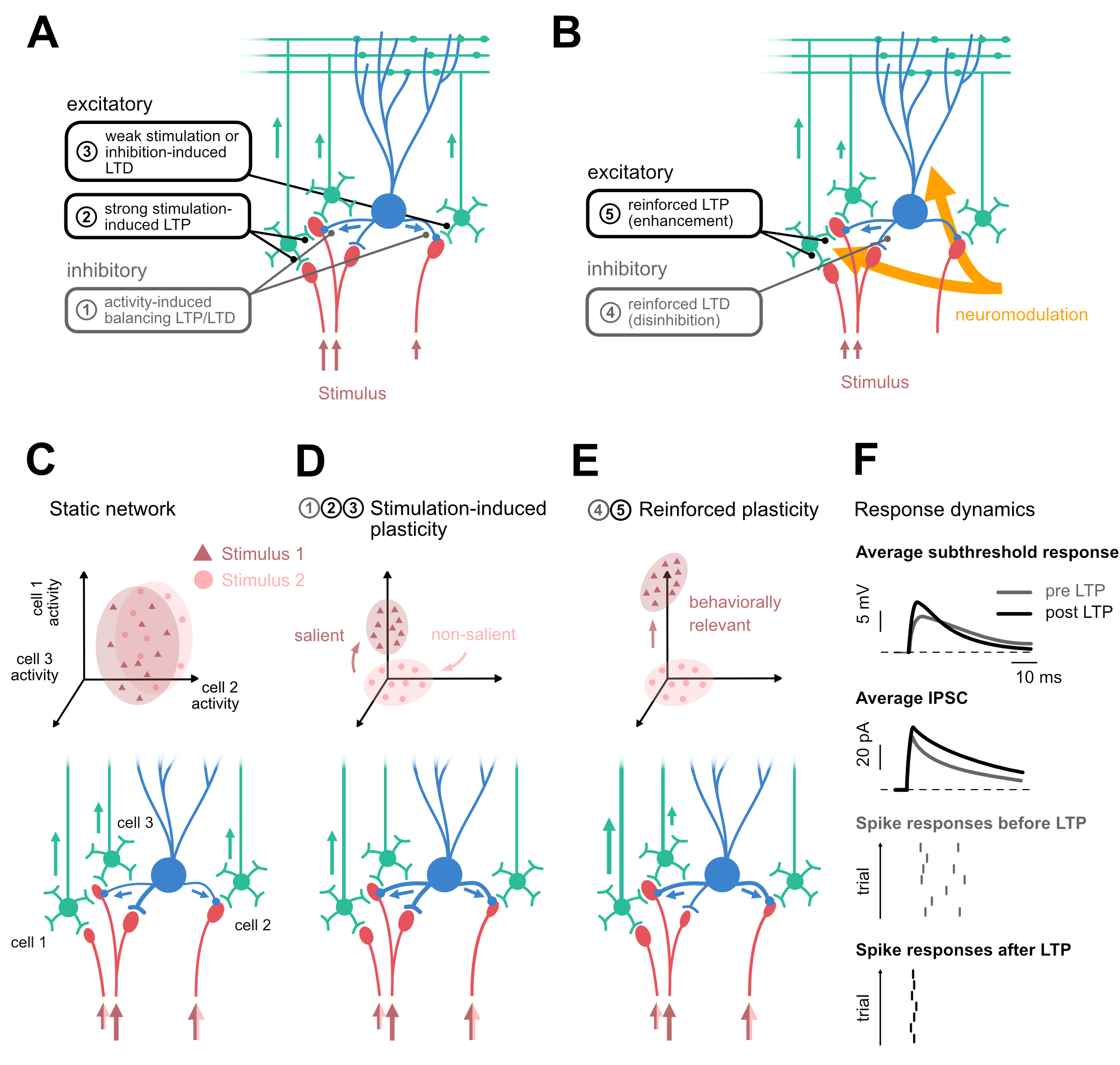} 
\caption{\textbf{Plasticity improves representation of salient and behaviorally relevant inputs.}
\textbf{(A)} Stimulation-induced plasticity of excitatory or inhibitory synapses adapt input representations in the expansion layer (main text). Inhibitory synapses undergo potentiation or depression to balance postsynaptic activation (1). For excitatory synapses, strong stimulation induces long-term potentiation (LTP) at dendrites with little inhibitory feedback and strong depolarization (2), while LTD is induced at dendrites experiencing weaker stimulation or stronger inhibition (3). 
\textbf{(B)} Reinforced plasticity occurs for coincident input activation and neuromodulatory signals and leads to disinhibition through LTD of inputs to interneurons (4), or direct enhancement through LTP of inputs to principal cells (5).
\textbf{(C--E)} Plasticity can adapt expansion layer responses to different stimuli to improve associative learning (main text).
\textbf{(C)} A static network might yield highly overlapping expansion layer activity for distinct stimuli (top), e.g., because they are activating partially overlapping inputs (bottom). 
\textbf{(D)} Stimulation-induced plasticity can improve expansion layer representations. Here, inhibitory plasticity reduces the overlap between responses by balancing redundant inputs. In parallel, excitatory plasticity sharpens the tuning of principal cells, where LTP enhances strong, salient stimuli and LTD suppresses weak or redundant ones.
\textbf{(E)} Reinforced inhibitory and excitatory plasticity enhances behaviorally relevant stimuli that coincide with neuromodulatory signals (here only Stimulus 1) to boost association strength and generalization (Box~\ref{info:theory}).
\textbf{(F)} After LTP, also the temporal response dynamics are altered. (Top) The average substhreshold response of principal cells becomes sharper after LTP (black) compared to before (grey) (c.f.,~\cite{mapelli2016heterosynaptic, Baltruschat2020}). 
(Middle) This faster decay can be caused by inhibitory plasticity that leads to stronger and longer-lasting IPSCs at the potentiated cells~\cite{mapelli2016heterosynaptic}.
(Bottom) Together, excitatory and inhibitory plasticity lead to a fast, precise, and efficient spike response~\cite{nieus2006ltp, mapelli2016heterosynaptic}.
}
\label{fig:plasticity}
\end{figure}

Taken together, stimulation-induced LTP and LTD enable a bidirectional modulation of inputs to expansion layers that could enhance strong, salient inputs and suppress weak, non-salient inputs (Figure~\ref{fig:plasticity}A,D), with a strong influence on the timing and reliability of responses~\cite{nieus2006ltp, d2009timing} (Figure~\ref{fig:plasticity} F). 
In addition, LTP and LTD induction are highly sensitive to inhibition, which could ensure non-redundant, sharply tuned input representations.

\subsection*{Reinforced plasticity of principal cell excitation}\label{sec:reinforcedPrinCell}

Reinforced plasticity of principal cell excitation refers to the facilitation, or long-lasting synaptic changes, enabled by neuromodulatory signals that encode behavioral relevance. 
In the mushroom body (MB) calyx, such plasticity is typically induced after sensory experience paired with appetitive or aversive stimuli, thus selectively reinforcing neural pathways associated with behaviorally relevant cues (see Table~\ref{tab:reinforced_principle_cells} for an overview). 
In the cerebellum granule layer, although generally less well explored, pharmacological application of nicotine---an agonist of acetylcholine receptors---has been shown to facilitate synaptic plasticity~\cite{prestori2013gating, cao2022nicotine}, demonstrating  reinforced plasticity in the presence of neuromodulatory attention-like signals. 
Strikingly, although sources of neuromodulation differ greatly between both systems, experimental findings point towards shared mechanisms through which reinforcement enhances representations of behaviorally relevant inputs.

In the MB calyx, reinforcement-driven plasticity has been demonstrated across species, where pairing odor stimuli with appetitive or aversive outcomes selectively modifies Kenyon cell responsiveness. 
These changes are most pronounced after long-term memory (\textbf{LTM}) formation through repeated spaced trainings.
For example, appetitive LTM in \textit{Drosophila} results in an increase in the number of microglomeruli responding to the conditioned odor and structural reorganization within these same synaptic complexes~\cite{Baltruschat2020}. 
This is paralleled by findings in honey bees and leaf-cutting ants, where appetitive and aversive LTM causes an increase in MG density~\cite{hourcade2010long, falibene2015long}, whereas the exposure to non-harmful stimuli decreases their density~\cite{falibene2015long}.
Similarly, aversive LTM formation in \textit{Drosophila} has been shown to lead to sequential increase in the activity of different KC subtypes within 9–48 hours~\cite{yu2006drosophila, akalal2010late}, and an increase in the number of responsive KCs to the conditioned odor~\cite{Delestro2020}.

In addition to long-term memory, appetitive, but not aversive,  short-term memory results in changes of Kenyon cell responses. Pairing an odor with sucrose for a few brief trials produces a robust increase in odor-evoked dendritic \calcium{} responses within at least 5–15 minutes in both, honey bees and \textit{Drosophila}~\cite{szyszka2008associative,louis2018cyclic}, revealing a rapid, physiological form of reinforcement. 
Thus, the MB calyx exhibits both rapid physiological and slow structural reinforcement mechanisms, organized across cell types and timescales to encode the behavioral relevance of experience.



Mechanistically, experimental evidence suggests that reinforcement in the MB calyx is driven by the coincidence of sensory input and neuromodulatory signals.  
Pairing stimulation of dopaminergic neurons with odor presentation enhances KC responses, whereas unpaired activation fails to induce plasticity, establishing a cellular substrate for reinforcement learning \cite{boto2019independent, qiao_input-timing-dependent_2022}. Because calcium signaling is crucial for stimulation-induced plasticity, reinforced plasticity likely arises through facilitation of \calcium{} transients both in bees and flies~\cite{perisse2009early, locatelli2005focal, muller2000prolonged}. Both dopamine and octopamine evoke \calcium{} influx via VGCCs in \textit{Drosophila}, yet only octopamine produces a synergistic increase when paired with cholinergic input~\cite{leyton2014octopamine}. Another key mechanism is cyclic adenosine monophosphate–protein kinase A (cAMP-PKA) activation~\cite{louis2018cyclic,fiala1999reversible}. Notably, appetitive—but not aversive—learning increases cAMP levels in the calyx~\cite{louis2018cyclic}, and octopamine, but not dopamine, strongly elevates cAMP~\cite{tomchik2009dynamics}. Both, the enhanced effect of octopamine on calcium and cAMP, are consistent with short-term enhancement of KC responses being restricted to appetitive conditioning~\cite{louis2018cyclic}. Together, these pathways link calcium and cAMP signaling to the selective reinforcement of behaviorally relevant inputs.

In cerebellum, although generally less explored, a similar reinforced plasticity as in the MB calyx has been observed through application of the neuromodulator nicotine. Both, in slice and in vivo, application of nicotine could shift the sign of plasticity from depression to potentiation via activation of $\alpha$7 nicotinic acetylcholine receptors~\cite{prestori2013gating, cao2022nicotine}. These receptors can facilitate \calcium-dependent plasticity through both, a direct calcium influx as well as depolarization of GrC dendrites~\cite{prestori2013gating}, thus facilitating further \calcium{} influx through VGCC and NMDARs. Moreover, if strongly activated, they might even on their own be sufficient to increase \calcium{} to a level that is typically reached through NMDA receptors. This is in line with the fact that reinforced plasticity even occurs when NMDARs are blocked~\cite{cao2022nicotine}. Thus, similar to observations in calyx, nicotinic acetylcholine receptors facilitate stimulation-induced plasticity to enhance responses for relevant afferent inputs.

In conclusion, neuromodulatory pathways reinforce short- and long-term plasticity in the calyx and granule layer by promoting depolarization and elevating intracellular \calcium{} and cAMP, highlighting shared mechanisms through which reinforcement enhances representations of behaviorally relevant inputs.

\subsection*{Plasticity of inhibitory interneurons}
Plasticity extends beyond excitatory synapses onto principal cells to include inhibitory circuits in both, the MB
calyx and the cerebellar granule layer. Because inhibitory interneurons integrate feedforward and feedback signals, changes at their synapses can reshape inhibitory responses and, consequently, sensory representations in expansion layers~\cite{fleming2024, pali2024understanding}. By potentiating inhibition for irrelevant inputs and promoting disinhibition for relevant ones, inhibitory plasticity could complement excitatory mechanisms to refine behaviorally meaningful coding.

In the cerebellum, plasticity of Golgi cells (GoCs) can adapt feedforward and feedback inhibition to patterns of input stimulation.
To adapt feedforward inhibition, high-frequency theta-burst stimulation induces both LTP and LTD at afferent MF–GoC synapses, as shown in cerebellar slices~\cite{locatelli2021calcium}.
Interestingly, however, the polarity of induction is reversed compared to the excitatory MF–GrC connections: 
LTP occurs at hyperpolarized potentials and depends on slow L-type VGCCs, whereas LTD emerges at depolarized potentials through T-type channels, independent of NMDARs~\cite{locatelli2021calcium}. 
A similar observation is made for feedback inhibition, where strong PF stimulation induces LTD at PF–GoC synapses~\cite{robberechts2010long}. 
Finally, MF and PF inputs drive timing-dependent plasticity in Golgi Cells~\cite{pali2025coincidence}, where dendritic integration across apical and basal compartments governs bidirectional synaptic changes and enables flexible inhibitory adaptation. 
Thus, stimulation-induced plasticity dynamically tunes both feedforward and feedback inhibition in the cerebellum.
Importantly, due to the apparent reversal in plasticity induction, this plasticity might complement excitatory plasticity through disinhibition or potentiation of inhibitory responses. 

An important question is therefore whether inhibitory interneurons also exhibit enhanced plasticity in the presence of reinforcement or relevance signals, and how those signals modulate the polarity of plasticity. Evidence suggests that interneuron responses are altered when sensory stimulation is paired with neuromodulatory input. In the cerebellum, suppression of Golgi cells by peripheral afferents decreases when these inputs coincide with climbing-fiber (\textbf{CF}) activation~\cite{xu2008climbing}, possibly due to transient inhibition of Golgi firing following CF stimulation. Similarly, in the MB calyx, inhibitory odor responses are reduced when odors are paired with reward in honey bees~\cite{grunewald1999physiological} and in the inhibitory APL neuron of \textit{Drosophila}~\cite{liu2009gabaergic}. Moreover, activating dopaminergic PPL2 neurons during odor presentation induces a long-term depression of APL responses via inhibitory Dop2R receptors~\cite{zhou_suppression_2019}. Overall, these results suggest that reinforcement signals can bias inhibitory plasticity toward disinhibition of behaviorally relevant inputs.

Plastic changes may also occur at inhibitory synapses onto principal cells, as proposed in~\cite{gao2012distributed}. 
Such inhibitory plasticity has been explicitly observed in the cerebellum granule layer, where the same induction cascade for LTP at MF-GrC synapses also drives the potentiation of the inhibitory synapse onto the same microglomerulus~\cite{mapelli2016heterosynaptic}. 
This is mediated by post-synaptic NMDARs and \calcium{} at GrC dendrites that then activates the retrograde messenger NO. 
In contrast, LTD induction only requires presynaptic NMDA receptors at GoC axon terminals. 
Consistent with the induction mechanisms, the expression of LTP and LTD has a strong presynaptic component in the form of altered synaptic release~\cite{mapelli2016heterosynaptic}. 
Functionally, inhibitory LTP causes a temporal sharpening of GrC responses in the presence of inhibition that, together with LTP at MF-GrC synapses, enables precise GrC responses although the total amount of depolarization decreases~\cite{mapelli2016heterosynaptic} (Figure~\ref{fig:plasticity}F). 
Therefore, heterosynaptic inhibitory plasticity complements excitatory plasticity at the parallel MF-GrC synapses as predicted by predictive learning theories (Box~\ref{info:theory}). 


In sum, plasticity of inhibitory synapses could ensure that inhibitory responses are connected to specifically those principal cell dendrite where they are most effective in suppressing inputs. 
Inhibitory responses, in turn, can be adapted by plasticity of afferent and feedback connections onto inhibitory interneurons. 
Here, experimental evidence suggests that, overall, inhibitory responses are weakened for salient or behaviorally relevant stimuli, whereas they are strengthened to irrelevant stimuli. 
This provides a complementary means of tuning representations in the expansion layer through a dynamic interplay of inhibitory and excitatory plasticity.


\begin{table}[h] \vspace{0.5ex}
\renewcommand{\arraystretch}{1.5}
\caption{Experimental observations of stimulation-induced potentiation of principal cells}\label{tab:stim_potentiation}
\footnotesize{
\begin{tabular}{>{\raggedright\arraybackslash}p{2.5cm}>{\raggedright\arraybackslash}p{2cm}>{\raggedright\arraybackslash}p{2cm}p{6cm}} 
\hline\hline 
\textbf{Reference} & \textbf{System} & \textbf{Experiment}  & \textbf{Observation} \\ [0.5ex] \hline 
\textcite{rossi1996differential} & Cerebellum granule layer & Acute rat slice & Simultaneous activation of mGluRs and NMDARs leads to potentiation of both NMDA and non‐NMDA receptor‐mediated synaptic transmission (especially slow component of NMDA transmission).\\
\textcite{d1999evidence} & Cerebellum granule layer & Acute rat slice   & High-frequency stimulation induces EPSC potentiation that depends on post-synaptic depolarization, \calcium{} and PKC, NMDAR and GluRs. Also observe NMDA current slow-down with potentiation. \\ 
\textcite{armano2000long} & Cerebellum granule layer & Acute rat slice & Increased excitability after high-frequency stimulus through increased input resistance and decreased spike threshold (dependent on depolarization, NMDARs and were prevented by inhibitory synaptic activity). Weaker inputs required than for potentiation of synaptic efficacy. \\ 
\textcite{maffei2002presynaptic} & Cerebellum granule layer & Acute rat slice & Presynaptic component of LTP after high-frequency stimulation (Ca2+, NMDAR and GluR dependent). \\  
\textcite{sola2004increased} & Cerebellum granule layer & Acute rat slice  & Increased neurotransmitter release can explain LTP at mossy fibre–granule cell synapses, as well as decreases in CV of EPSPs, failure of release and PPR. Strong evidence that LTP is expressed presynaptically. \\ 
\textcite{gall2005intracellular} & Cerebellum granule layer & Acute rat slice & Record \calcium{} dependence curve of LTP similar to BCM rule (as well as NMDAR and GLuR dependence). \\  
\textcite{nieus2006ltp} & Cerebellum granule layer & Acute rat slice & Potentiation through high-frequency stimulation, record and model presynaptic (increase of release) and postsynaptic contributions (conductance) that shift GC burst initiation and frequency, respectively. Increased AMPA and NMDA currents. \\  
\textcite{roggeri2008tactile} & Cerebellum granule layer & In vivo, anesthetized rat  & Predominantly MF-GrC LTP after $\qty{20}{\hertz}$ facial whisker stimulation with inhibition blocked (gabazine), else LTD.\\  
\textcite{d2009differential} & Cerebellum granule layer & Acute rat slice  & Frequency-dependence of LTP, find that \calcium-dependence underlies both duration and frequency-dependence of LTD/LTP. \\  
\textcite{sgritta2017hebbian} & Cerebellum granule layer & Acute rat slice &  EPSP-spike pairing at 6 Hz optimally induces STDP at MF-GC synapse in rats (LTP and LTD were confined to a ±25 ms time-window). STDP  vanished $>50\,\text{Hz}$ (only LTP ) or $<1\,\text{Hz}$ (only LTD). Its sign is inverted by GABA$_\text{A}$ activaton. \\ 
\textcite{lu2022facial} & Cerebellum granule layer & In vivo, anesthetized mice  & MF-GrC LTP after 20 Hz facial stimulation and blocking of GABA$_\text{A}$ receptors via the GluN2A-containing NMDA receptor/nitric oxide cascade in mice (presynaptic expression). \\ 
 \textcite{qiao_input-timing-dependent_2022} & MB calyx & Drosophila, in vivo & Timing-dependent potentiation of PN-$\gamma$ KC synapses after pre-post pairing, which depends on D2-like dopamine receptors. 
 \\ 
\hline\hline
\end{tabular}}
\end{table}
\normalsize

\begin{table}[h] \vspace{0.5ex}
\renewcommand{\arraystretch}{1.5}
\caption{Experimental observations of stimulation-induced depression of principal cells}\label{tab:stim_depression}
\footnotesize{
\begin{tabular}{>{\raggedright\arraybackslash}p{2.5cm}>{\raggedright\arraybackslash}p{2cm}>{\raggedright\arraybackslash}p{2cm}p{6cm}} 
\hline\hline 
\textbf{Reference} & \textbf{System} & \textbf{Experiment} & \textbf{Observation} \\ [0.5ex] \hline
\textcite{gall2005intracellular} & Cerebellum granule layer & Acute rat slices & MF-GrC LTD after stimulation with short pulses (\qty{100}{\milli\second} at \qty{100}{\hertz}). NMDAR dependent.\\ 
\textcite{mapelli2007spatial} & Cerebellum granule layer & Acute rat slices & MF-GrC LTD after TBS, with LTD prevailing in regions with negative E-I balance.
\\  
\textcite{roggeri2008tactile} & Cerebellum granule layer & In vivo, anesthetized rat & LTP and LTD in after facial tactile and intracerebellar electrical stimulation. LTD prevailed in control conditions, whereas LTP prevailed during local application of gabazine (in line with \cite{mapelli2007spatial}).
\\ 
\textcite{d2009differential} & Cerebellum granule layer & Acute rat slices & MF-GrC LTD for slow frequencies. Similar \calcium-dependence underlies both duration and frequency-dependence of LTD/LTP, NMDAR-dependent.(cf.~\cite{gall2005intracellular}). STP changes reverse under LTD (increase in PPR and CV, reduced reliability). NMDAR-dependent. 
\\  
\textcite{sgritta2017hebbian} & Cerebellum granule layer & Acute rat slice  &  EPSP-spike pairing at 6 Hz optimally induces STDP at MF-GC synapse in rats (LTP and LTD were confined to a ±25 ms time-window). STDP  vanished $>50\,\text{Hz}$ (only LTP ) or $<1\,\text{Hz}$ (only LTD). Its sign is inverted by GABA$_\text{A}$ activaton. \\ 
\textcite{casali2020cellular} & Cerebellum granule layer & Acute rat slice   & \calcium{} imaging confirms center-surround organization of LTP/LTD similar to~\cite{mapelli2007spatial}.\\ 
\textcite{lu2022facial} & Cerebellum granule layer & In vivo, anesthetized mice   & MF-GrC LTD after 20 Hz facial stimulation, NMDAR-dependent. \\   \textcite{qiao_input-timing-dependent_2022} & MB calyx & Drosophila, in vivo & Timing-dependent potentiation of PN-$\gamma$ KC synapses after pre-post pairing, which depends on D2-like dopamine receptors.
 \\ 
\textcite{szyszka2008associative} & MB calyx & In vivo, honey bee  & Progressive reduction in KC responses upon repeated odor exposure that lasts for at least 5 minutes.
\\  
\textcite{louis2018cyclic} & MB calyx & In vivo, Drosophila  & Depression of KC responses after repeated odor presentation (saline control) that lasts for at least 5 minutes. 
\\ 
\hline\hline
\end{tabular}}
\end{table}
\normalsize

\begin{table}[h] \vspace{0.5ex}
\renewcommand{\arraystretch}{1.5}
\caption{Overview of reinforced plasticity at principal cells.}\label{tab:reinforced_principle_cells}
\footnotesize{
\begin{tabular}{>{\raggedright\arraybackslash}p{2.5cm}>{\raggedright\arraybackslash}p{2cm}>{\raggedright\arraybackslash}p{2cm}p{6cm}} 
\hline\hline 
\textbf{Reference} & \textbf{System} & \textbf{Experiment} & \textbf{Observation} \\  [0.5ex] \hline
\textcite{Delestro2020} & MB calyx & In vivo, Drosophila &  After long-term memory (LTM) formation number of responsive KC somata increases and single neuron signal become stable. \\ 
\textcite{hourcade2010long} & MB calyx & In vivo, honey bee & Density of microglomeruli increases after appetitive LTM formation. \\
\textcite{falibene2015long} & MB calyx & In vivo, leaf-cutting ants & Density of microglomeruli increases after aversive LTM formation, but decreases after prolonged exposure to neutral stimuli. \\
\textcite{Baltruschat2020} & MB calyx & In vivo, Drosophila &  Number and size of microglomeruli that respond to the conditioned odor increase after appetitive LTM formation. \\ 
\textcite{szyszka2008associative} & MB calyx & In vivo, honey bee  & Increase of activity in specific KC dendrites after appetitive olfactory learning. \\ 
\textcite{qiao_input-timing-dependent_2022} & MB calyx & In vivo, Drosophila & PPL1 activation (dopaminergic feedback) facilitates stimulation-induced potentiation of PN-$\gamma$ KC synapses. \\ 
\textcite{boto2019independent} & MB calyx & In vivo, Drosophila   & Reinforcement of activity in KCs through PPL2 innervation, increases memory strength independent of valence \\ 
\textcite{louis2018cyclic} & MB calyx & In vivo, Drosophila &  Elevating cAMP drives increases in KC odor responses (most strongly in $\gamma$KCs. cAMP levels are increased by appetitive learning or activation of DANs. Appetitive, but \emph{not} aversive learning,  shows compensation of depression similar to \textcite{szyszka2008associative}. 
\\ 
\textcite{yu2006drosophila} & MB calyx & In vivo, Drosophila  & Long-term increase in axonal \calcium{} in $\alpha/\beta$-KCs after multiple, spaced odor-shock pairings. 
\\ 
\textcite{akalal2010late} & MB calyx & In vivo, Drosophila &  Long-term increase in axonal \calcium{} in $\gamma$-KCs after multiple, spaced odor-shock pairings. 
\\ 
\textcite{prestori2013gating} & Cerebellum granule layer & Acute rat slice   & Activation of $\alpha$7 nicotinic acetylcholine receptors ($\alpha$7nAchRs) on MF terminals and GC dendrites enhances postsynaptic \calcium{} influx, which can turn LTD into LTP (NMDAR dependent?). 
\\ 
\textcite{cao2022nicotine} & Cerebellum granule layer & In vivo, anesthetized mice & Facial 20 Hz stimulation combined with nicotine showed enhanced MF-GrC LTP via the $\alpha$4$\beta$2 nAChR/NO signaling pathway (NMDA independent).
\\
\hline\hline
\end{tabular}}
\end{table}
\normalsize

\begin{table}[h] \vspace{0.5ex}
\renewcommand{\arraystretch}{1.5}
\caption{Experimental observations of stimulation-induced plasticity at inhibitory interneurons}\label{tab:stim_inhibition}
\footnotesize{
\begin{tabular}{>{\raggedright\arraybackslash}p{2.5cm}>{\raggedright\arraybackslash}p{2cm}>{\raggedright\arraybackslash}p{2cm}p{6cm}} 
\hline\hline 
\textbf{Reference} & \textbf{System} & \textbf{Experiment} & \textbf{Observation} \\  [0.5ex] \hline
 \textcite{robberechts2010long} & Cerebellum granule layer & acute rat slices   & LTD of PF to Golgi cell synapses (anti-Hebbian plasticity) after strong PF input 
 \\ 
 \textcite{ruediger2011learning} & Cerebellum granule layer & mouse behav. experiments + morphological analyses & Learning-related growth of feedforward inhibition (MF to Golgi cells) in specific lobuli (could also be reinforced).
 \\  
\textcite{locatelli2021calcium} & Cerebellum granule layer & Acute rat slices   & MF theta-burst stimulation (TBS) induced either LTP or LTD at MF-Golgi cell and GC-Golgi cell synapses. LTD or LTP being favored when TBS induction occurred at depolarized or hyperpolarized potentials, respectively. Thus, voltage dependence of plasticity at MF-Golgi cell  synapses was inverted w.r.t plasticity at MF-GC synapse. 
\\ 
\textcite{pali2025coincidence} & Cerebellum granule layer &  Acute rat slices   & Hebbian STDP at MF-GoC synapse under pairing of MF input and PF induced APs at the GoC at theta frequencies (4-6 Hz). STDP curve is inverted when unblocking GABA$_\text{A}$ receptors. NMDAR-dependent and most likely postsynaptic expression. Pairing MF with MF-induced AP induces LTD with pre or post pairing, pairing MF with stimulated AP induces no plasticity.\\
\hline\hline
\end{tabular}}
\end{table}
\normalsize


\begin{table}[h] \vspace{0.5ex}
\renewcommand{\arraystretch}{1.5}
\caption{Overview of reinforced plasticity at inhibitory interneurons.}\label{tab:reinforced_inhibition}
\footnotesize{
\begin{tabular}{>{\raggedright\arraybackslash}p{2.5cm}>{\raggedright\arraybackslash}p{2cm}>{\raggedright\arraybackslash}p{2cm}p{6cm}} 
\hline\hline 
\textbf{Reference} & \textbf{System} & \textbf{Experiment} & \textbf{Observation} \\  [0.5ex] \hline
 \textcite{grunewald1999physiological} & MB calyx & In vivo, honey bee  & GABAergic feedback neurons in the mushroom body show responses to odor and sucrose stimuli in intracellular recordings. After single paired presentation of odor and sucrose, interneurons show decreased odor responses briefly after conditioning, and increased responses for either odor or sucrose stimulation.\\  
 \textcite{liu2009gabaergic} & MB calyx & In vivo, Drosophila  & APL responds to both, odor and shock stimuli. APL responses are suppressed by pairing an odor stimulus with 12 shocks, with conditioned depression being present already $\qty{30}{\sec}$, but stronger depression is observed at $\qty{5}{\min}$. \\ 
\textcite{haenicke2018neural} & MB calyx & honey bee   & Temporal profile of induced changes of odors in calyx after appetitive conditioning matches that of GABAergic PCT neurons.
\\ 
\textcite{zhou_suppression_2019} & MB calyx & In vivo, Drosophila   & APL is depressed through direct PPL1 input and DD2Rs, yielding a disinhibition and also subsequent suppression after conditioning. Knockdown of either DD2R or its downstream molecules in the APL neurons impairs olfactory learning at the behavioral level.
\\ 
\textcite{ruediger2011learning} & Cerebellum granule layer & mouse behav. experiments + morphological analyses & Learning-related growth of feedforward inhibition (MF to Golgi cells) in specific lobuli (could also be reinforced).\\  
\textcite{xu2008climbing} & Cerebellum granule layer & In vivo, anesthetized rat   & Golgi cells are depressed by climbing fiber inputs (probably indirectly), providing reinforced disinhibition. Long-term changes of afferent responses in Golgi cells upon pairing CF inputs with afferent inputs (reduction of Golgi cell suppression by peripheral inputs).\\ 
\hline\hline
\end{tabular}}
\end{table}
\normalsize


\section*{Functional relevance of plasticity in expansion layers}

Ample experimental evidence indicates that expansion layers in cerebellum-like structures are subject to plasticity. However, the functional contribution of these modifications to adaptive behavior, particularly, how they shape sensory representations to support associative learning remains an open question.
Here, we propose two complementary roles for plasticity in expansion layers that are based on experimental evidence from both, the mushroom body and cerebellum, as well as recent modeling studies and theories of representation learning.


First, we propose that stimulation-induced plasticity, which occurs in the absence of explicit reinforcement, enables expansion layers to adapt to features of their inputs—thus improving future associative learning.
Second, we propose that reinforced plasticity within expansion layers is complementary to  associative learning at the output layer by encoding association strength and boosting generalization.
In this view, we argue that plasticity in expansion layers and the output stage reflects a normative division of labor, where the former extracts salient input features, while the latter associates these features with specific adaptations in behavior.


\begin{figure}
\begin{infobox}[label={info:theory}]{Theories of unsupervised representation learning}
The core principle of many unsupervised representation learning theories is to re-code (sensory) data into a more amenable form while preserving information~\cite{olshausen1997sparse,lee2000unifying}. By doing so, these models can identify the latent factors, or \textit{causes}, of sensory data~\cite{olshausen1997sparse}, which can improve behavioral learning~\cite{gershman2015discovering}. 
Importantly, most models rely on a similar architecture as the expansion layer of cerebellum-like structures, where principal neurons encode inputs from sensory neurons and compete via lateral inhibition~\cite{foldiak1990forming,linsker1997local,schweighofer2001unsupervised,brendel2020learning}.
However, these models differ in their learning rules. 
\paragraph{Competitive Hebbian learning} According to Hebbian learning,
principal neurons with non-zero firing rate $r^{\text{(enc)}}_j>0$ change their input weights $W^{\text{(inp)}}_{ji}$ to encode the current input pattern
\begin{equation*}
    \Delta W^{\text{(inp)}}_{ji} \propto r^{\text{(enc)}}_j(r^{\text{(inp)}}_i -  W^{\text{(inp)}}_{ji}) ,\quad\quad\text{(Hebbian learning)}
\end{equation*}
where $r^{\text{(inp)}}_i$ are the firing rates of the input neurons.
This is a form of the classic Oja's rule, which has been applied in many models before~\cite{mikulasch2021local}.
In addition, inhibitory connections are often adjusted via anti-Hebbian plasticity to enhance competition between similarly tuned principal neurons and enforce independent firing~\cite{foldiak1990forming,schweighofer2001unsupervised,vogels2011inhibitory}. Since these rules only depend on supra-threshold spiking activity, these models use point neurons.

 \vspace{0.5em}
 \begin{minipage}[t]{.48\textwidth}
\textbf{Function:}\\
Through the interplay of lateral inhibition and excitatory plasticity, neurons learn to identify independent factors of the data \cite{foldiak1990forming}.

\vspace{0.5em}
\textbf{Limitations:}\\
Leads to bad solutions with realistic transmission delays and correlated input features~\cite{mikulasch2021local}.
\end{minipage}\hfill
\begin{minipage}[t]{.48\textwidth}
\textbf{Observed features:}
 \vspace{-0.6em}
\begin{itemize}
    \item  Coincident input and post-synaptic depolarization lead to potentiation or depression, depending on the input strength (Table~\ref{tab:stim_potentiation} and~\ref{tab:stim_depression}).
    \vspace{-0.5em}
    \item Inputs to a single principal neuron are more correlated than random inputs~\cite{gruntman2013integration}.
\end{itemize}
\end{minipage}

\paragraph{Dendritic predictive learning}
In predictive learning, the goal is to predict sensory inputs from responses in the network.
To this end, inhibitory connections $W^{\text{(inh)}}_{ik}$ learn to cancel predictable (i.e., redundant) sensory inputs~\cite{deneve2016efficient, brendel2020learning, mikulasch2021local}, while input weights $W^{\text{(inp)}}_{ji}$ adapt to encode the unpredicted, salient inputs~\cite{linsker1997local,coenen2001parallel,mikulasch2021local} via 
\begin{align*}
    \Delta W^{\text{(inh)}}_{ij} &\propto r^{\text{(inh)}}_j \epsilon_i  ,\quad\quad\text{(Inhibitory balance)}\\
        \Delta W^{\text{(inp)}}_{ji} &\propto r^{\text{(enc)}}_j \epsilon_i .\quad\quad\text{(Error-based learning)}
\end{align*}
Here, $\epsilon_i=r^{\text{(inp)}}_i - \sum_k W^{\text{(inh)}}_{ik} r^{\text{(inh)}}_k$ is the prediction error between the presynaptic firing $r^{\text{(inp)}}_i$ of input $i$ and the internal prediction that is based on inhibitory responses $r^{\text{(inh)}}_k$.
Theoretical work has shown that this prediction error can be locally computed by electrically decoupled dendritic compartments that integrate excitatory and inhibitory inputs~\cite{mikulasch2021local}, similar to 
dendrites of principal neurons in the expansion layer~\cite{kanichay2008synaptic, leiss2009synaptic, gruntman2013integration}. 

 \vspace{0.5em}
 \begin{minipage}[t]{.48\textwidth}
\textbf{Function:}\\
Same as Hebbian learning, yet heterosynaptic excitatory and inhibitory plasticity coordinate learning in the network to yield non-redundant input representations even for correlated firing or inhibitory transmission delays~\cite{mikulasch2021local}.

\vspace{0.5em}
\textbf{Limitations:}\\
Requires many inhibitory synapses and connections, in principle between every principal neuron to each dendrite. However, microglomeruli provide a very efficient wiring scheme because one inhibitory connection can target multiple dendrites simultaneously.
\end{minipage}\hfill
\begin{minipage}[t]{.48\textwidth}
 \textbf{Additional observed features:}
 \vspace{-0.5em}
 \begin{itemize}
     \item Dendritic inhibition biases excitatory plasticity towards LTD~\cite{mapelli2007spatial, casali2020cellular, lu2022facial}.
     \vspace{-0.2em}
     \item Inhibition is also plastic and depends on the same induction cascade as excitatory plasticity (same error dependence)~\cite{mapelli2016heterosynaptic}.
     \vspace{-0.2em}
     \item Due to inhibitory plasticity, microglomeruli with more excitatory synapses also receive more inhibitory synapses~\cite{eichler2017complete, abdelrahman2021compensatory}, and  the magnitude of EPSCs and IPSCs is highly correlated at individual principal neurons~\cite{mapelli2016heterosynaptic}.
 \end{itemize}
\end{minipage}

\end{infobox}
\end{figure}

\subsection*{Non-associative plasticity supports generalization and discrimination}
In natural environments, animals must rapidly learn to adapt behavior from sparse reward associations to a broad array of unreinforced, yet behaviorally relevant stimuli.
In particular, behaviorally equivalent input contexts should robustly elicit the same behavior, despite perceivable variations in the inputs, termed \emph{generalization}, while different input contexts should not elicit the same behavior, despite perceivable commonalities, termed \emph{discrimination}~\cite{mishra2010adaptive}.
Both aspects are enhanced by compression in the input layer~\cite{muscinelli2023optimal} and pattern separation in the expansion layer as a consequence of circuit design and inhibitory feedback~\cite{cayco2019re, eichler2017complete, litwin2017optimal}. 
Here, we argue that expansion layer plasticity also contributes to generalization and discrimination of adaptive behavior by adapting input representations. 

In order to achieve good generalization, an efficient strategy in reinforcement learning is to first learn internal representations that capture the statistics of the environment, and to then form reward associations to these statistical features~\cite{moerland2023model, gershman2015discovering}.
This way, behavioral responses generalize to unseen stimuli that agree with the learned model of the environment, even if reward signals are very sparse.
Expansion layers in cerebellum-like circuits are ideally positioned in the information processing hierarchy to support such representation learning. But through what cellular mechanisms could such adaptive changes arise? 

Theoretically, learning input features can emerge from Hebbian and predictive learning~\cite{foldiak1990forming,brendel2020learning, mikulasch2021local}. These rules closely parallel experimentally observed stimulation‑induced LTP and LTD of input synapses in granule cells and Kenyon cells (see Box~\ref{info:theory} and Figure~\ref{fig:plasticity}A).
Consistent with this idea, a previous modelling study showed that Hebbian-like learning in the expansion layer decreases the error in subsequent associative or motor learning tasks~\cite{litwin2017optimal, schweighofer2001unsupervised}.
Moreover, experimentally constrained circuit models of the granule layer showed that synaptic plasticity of mossy fiber inputs adapts spatio-temporal filtering properties of granule cells to features of their inputs (e.g., input frequency)~\cite{nieus2006ltp,diwakar2011local, casali2020cellular}, yielding fine-tuned filters that can be used for downstream learning in the molecular layer~\cite{fujita1982adaptive}. 
Together, these results suggest that stimulation‑induced plasticity of input synapses in expansion layers can enable the learning of input features (Figure~\ref{fig:plasticity}E), thereby improving the efficiency and generalization of subsequent associative learning. 

In addition to variable inputs, cerebellum-like structures face the challenge that in natural environments many input features can be present in parallel, but only few are behaviorally relevant. 
To mitigate this, a key strategy is to suppress responses to nuisance background stimuli that have been previously experienced but did not coincide with any signal of behavioral relevance~\cite{pirez2023experience}. 
Such suppression, also termed \emph{habituation}, is well described in the insect antennal lobe, the input stage of the MB calyx, after long stimulation with an odor, where it biases responses away from habituated and towards novel stimuli~\cite{das2011plasticity, locatelli2013nonassociative}.
An even more rapid suppression of responses has also been reported in the MB calyx after few stimulus repetitions in honey bees and \emph{Drosophila}~\cite{szyszka2008associative, louis2018cyclic}, as well as after nicotine stimulation of cultured KCs~\cite{campusano2007nachr} or \textit{ex-vivo} recordings of \emph{Drosophila} after electrical stimulation of the antennal lobe~\cite{sato2018synaptic}.
Importantly, this rapid suppression is not found at the level of projection neurons~\cite{szyszka2008associative}, depends on \emph{rutagaba}, cAMP and calcium channels in KCs~\cite{campusano2007nachr,sato2018synaptic} and can last several minutes up to hours, suggesting that plasticity in the expansion layer is a likely cause for this suppression. 
Suppression of responses after repeated stimulation has also been observed in the granule layer as a consequence of a potentiation of inhibitory synapses~\cite{mapelli2016heterosynaptic} or long-term depression of afferent synapses~\cite{gall2005intracellular, mapelli2007spatial, d2009timing}.
This synaptic depression has been shown to contribute to the sharpening of granule cell tuning, thus improving pattern separation for overlapping inputs~\cite{casali2020cellular}. 
Therefore, excitatory and inhibitory plasticity in expansion layers might enable a rapid suppression of irrelevant background stimuli and overlapping inputs to improve discrimination of subsequently learned associations (Figure~\ref{fig:plasticity}D,E).

In sum, non‑associative, Hebbian‑like plasticity and repetition‑induced suppression could work together to enhance later associative learning. The former makes the circuit respond reliably to variable inputs, thereby improving generalization, while the latter suppresses responses to irrelevant stimuli and overlapping inputs, thus enhancing discrimination. However, this raises an apparent paradox: repeated exposure to a stimulus should both, increase reliability of responses that encode its features (e.g., through Hebbian potentiation), and weaken overall responses to this stimulus (suppression).
Strikingly, recent developments in predictive representation learning provide a possible solution to this through a cooperation of inhibitory and excitatory plasticity~\cite{mikulasch2021local}. Here, inhibition is rapidly potentiated to suppress predictable inputs, as observed in the granule layer~\cite{mapelli2016heterosynaptic}, and then gates synaptic potentiation at principal cell dendrites to allow enhanced responses only for a very selective set of principal cells (Box~\ref{info:theory} and Figure~\ref{fig:plasticity}A,E). 
Notably, a realistic model of cerebellum plasticity has also shown that the gating of MF plasticity by inhibition and post-synaptic calcium leads to a specialization of granule cells to different input features~\cite{casali2020cellular}.
Finally, the potentiation of excitation \emph{and} inhibition leads also to a temporal sparsening of responses over time~\cite{mapelli2016heterosynaptic}, hence enabling a low latency, temporally precise response instead of an overall increase of depolarization or spike rate (Figure~\ref{fig:plasticity}F)~\cite{d2009timing, pali2024understanding}.
Thus, a tight coordination of excitatory and inhibitory plasticity could be a mechanism that allows expansion layer plasticity to improve both, generalization and discrimination, as well as response timing for better pattern separation and subsequent associative learning.

\begin{figure}
\begin{infobox}[label={info:assoc}]{Encoding of association strength in expansion layers}
\begin{center}
    \includegraphics[width=\textwidth]{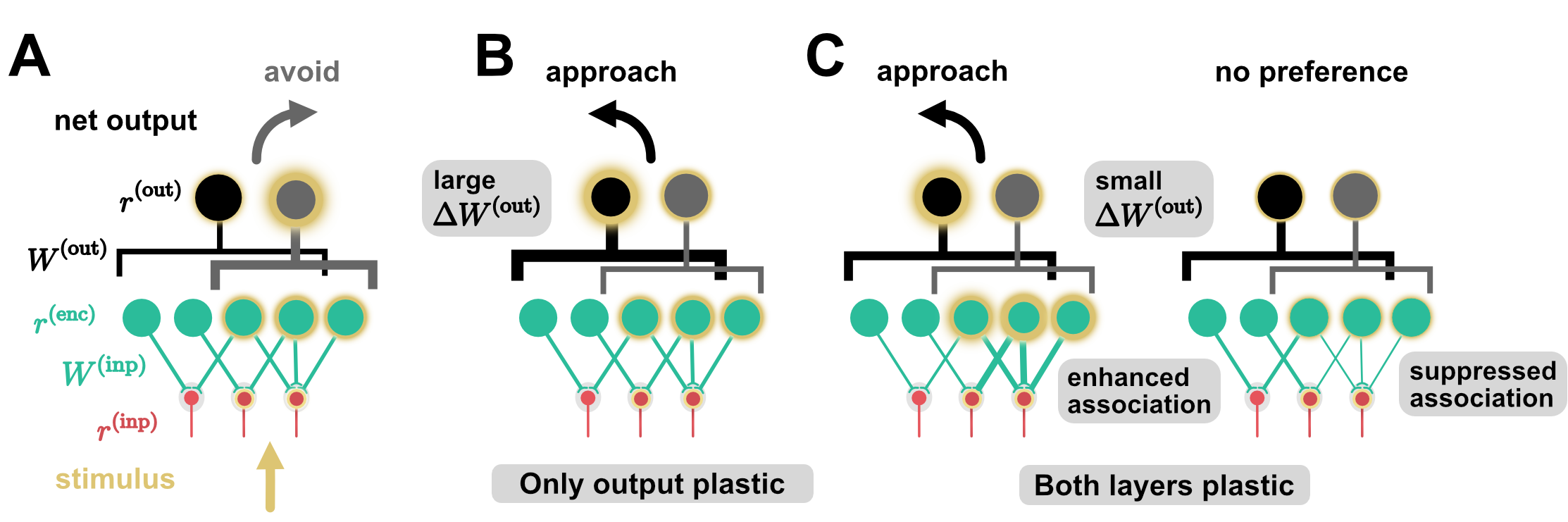}
\end{center}
Originally, it was thought that mainly synapses to the output neurons undergo reinforced plasticity to form associations in cerebellum-like circuits. However, recent evidence suggest that also reinforced plasticity of expansion layer responses contributes to associative learning. 
Here, we elaborate that expansion layer plasticity can encode the \emph{association strength} of a stimulus, which yields a division of labor that boosts flexibility and generalization of associative learning. 

Formally, the total behavioral output for a stimulus can be expressed as the product of expansion-layer activity \( r^{\text{(enc)}} \) and downstream synaptic weights \( W^{\text{(out)}} \):
\[
r^{\text{(out)}} = W^{\text{(out)}}\, r^{\text{(enc)}}.
\]
In this simplified formulation the output responses \( r^{\text{(out)}} \) are a weighted sum of their inputs and could, e.g., encode opposing behavioral modulation such as approach and avoidance (A,B). Thus, while changes in output weights \( W^{\text{(out)}} \) can invert the net behavioral output by differentially potentiating or depressing the different output neurons (from A to B), expansion layer responses \( r^{\text{(enc)}} \) modulate the \textit{strength} of the associated behavioral output (C). 
Although this model is a vast simplification of the intricate dynamics at the output stage, it accounts for the fact that the transmission between a principal cell in the expansion layer and an output neuron is a product of response and synaptic weight.
Importantly, while a fixed stimulus encoding requires substantial synaptic weight changes $\Delta W^{\text{(out)}}$ to achieve an adaptive behavior (from A to B), increasing responses for salient stimuli subsequently requires only small weight changes to achieve the same output (C, left). This has the following functional advantages:
\begin{itemize}
    \item \textbf{Flexible re- and reversal learning:} Enhancing expansion layer responses to a stimulus over prolonged timescales (e.g., through long-term potentiation of input synapses) enables small weight changes $\Delta W^{\text{(out)}}$ to rapidly restore strong adaptive behavior in re-learning or to reverse it during reversal learning. 
    \item \textbf{Suppression of learning for irrelevant, confounding stimuli:} Strengthening inhibitory synapses or weakening excitatory inputs can reduce responses to irrelevant stimuli, preventing these signals from driving adaptive behavioral changes despite ongoing synaptic plasticity (C, right).
    \item \textbf{Enhanced generalization:} By enhancing expansion layer responses (e.g., through potentiation of input synapses or depression of inhibition) not only is the association strength increased, but also do similar inputs elicit similar responses~\cite{peng2017simple}, thus enhancing generalization for variable inputs.
\end{itemize}
Therefore, a separate encoding of association strength of a stimulus in the expansion layer serves complementary functions to downstream plasticity.

\end{infobox}
\end{figure}

\subsection*{Reinforced plasticity enhances association strength and generalization}
While classical models place reinforcement-driven plasticity at the output of cerebellum-like structures, evidence presented above suggests that the expansion layers themselves also undergo associative modification. Pairing a sensory stimulus with neuromodulatory activation changes sensory representations in both the mushroom body (MB) calyx~\cite{yu2006drosophila, szyszka2008associative, akalal2010late, louis2018cyclic, ,zhou_suppression_2019, Delestro2020, Baltruschat2020,  qiao_input-timing-dependent_2022} and the cerebellar granule layer~\cite{prestori2013gating, cao2022nicotine}. 
This second, “input-level” stage of associative learning may not simply duplicate output plasticity but instead serve distinct and complementary functions for adaptive behavior.
This is in line with experimental work suggesting that associative learning in the MB calyx serves a separate encoding of \emph{association strength} of a stimulus, independent of the valence or adaptive behavior that is associated through learning at the output stage~\cite{boto2019independent}. 
Moreover, mossy fiber inputs that convey feedback signals from the cerebellar nucleus to the granule layer have been found to undergo plasticity and enhance the association strength in eyeblink conditioning in mice~\cite{gao2016excitatory}.
Interestingly, incorporating a separate learning of the association strength of a stimulus is also a key element in models of human associative learning~\cite{mackintosh1975theory, le2016attention, radulescu2019holistic}. 
Here, we argue that a separate learning of association strength in the expansion layer is also a general property of cerebellum-like structures (see Box~\ref{info:assoc} for a formal argument) and yields the following functional advantages. 

First, encoding association strength in the expansion layer enables robust adaptive behavior that also preserves the flexibility of future learning. 
This is because the association strength of a stimulus can be stored for long-term in the expansion layer without affecting the flexibility to associate a different behavior or valence through learning at the output synapse (Box~\ref{info:assoc}). 
This renders expansion layer plasticity an ideal complementary mechanism for long-term memory (LTM) and consolidation of adaptive behavior.
Consistent with this idea, \textit{Drosophila} studies show that Dop2R-dependent potentiation at PN–KC synapses in the expansion layer is required for aversive long-term but not short-term memory~\cite{qiao_input-timing-dependent_2022}, and that LTM formation induces input-specific microglomerular remodeling and enhanced KC reliability~\cite{Baltruschat2020, Delestro2020}.
This is also consistent with observations in leaf-cutting ants and honey bees, where long-term memory formation during avoidance and appetitive learning leads to an increase in the density of microglomeruli~\cite{falibene2015long, hourcade2010long, }, whereas prolonged exposure to neutral stimuli reduces their density~\cite{falibene2015long}.
In the cerebellum, long-term vestibulo-cerebellar learning similarly depends on NMDA receptor–mediated plasticity at mossy fiber–granule cell synapses~\cite{andreescu2011nr2a, seja2012raising}, and structural changes at mossy fiber-Golgi cell connections accompany cued fear conditioning and long-term motor learning~\cite{ruediger2011learning}.
Together, these findings suggest that plasticity in expansion layers provides a stable substrate for consolidation across species, enabling adaptive behavior to persist over long timescales while retaining flexibility for new learning.

Second, encoding association strength in the expansion layer may enhance the generalization of learned associations (see Box~\ref{info:assoc}).
This is in line with modelling studies showing that strengthening input synapses in the expansion layer through neuromodulatory signals enhances generalization in olfactory learning~\cite{peng2017simple} and reduces errors in a motor learning task~\cite{schweighofer2001unsupervised}.
However, this might also come at the cost of reduced discrimination~\cite{peng2017simple}, in line with experimental observations when stimulating octopamine release during olfactory learning in \emph{Drosophila}~\cite{wong_octopaminergic_2021}.
Intuitively, learning association strength in the expansion layer should thus be more pronounced for adaptive behavior that should generalize well (e.g., appetitive learning) compared to adaptive behavior that requires high specificity (e.g., aversive learning).
This is supported by work in \emph{Drosophila} that 
indicates a stronger effect of (appetitive) octopaminergic neuromodulation on calcium and cAMP increase in the calyx as compared to (aversive) dopamine, whereas octopamine has a relatively weaker effect on cAMP elevation in KC axons~\cite{tomchik2009dynamics, leyton2014octopamine}.
Moreover, response changes in \textit{Drosophila} calyx have been observed for appetitive, but not aversive short-term memory~\cite{louis2018cyclic}, and memory traces in PN-KC synapses have been shown to be sufficient for appetitive but not for aversive memory~\cite{thum2007multiple}.
Together, these results suggest a differential role of associative plasticity in expansion layers for appetitive and aversive learning, potentially through different requirements on generalization and specificity.

In sum, a separate encoding of association strength within the expansion layer may enhance both the flexibility and generalization of adaptive behavior.
Moreover, these same properties make expansion-layer plasticity a strategic target for neuromodulators that convey salience and attention signals, such as acetylcholine or glutamate~\cite{schweighofer2004cerebellar, ueno2013long, ueno2017coincident}. 
Increasing evidence indicates that such modulators facilitate potentiation of input synapses in the expansion layer~\cite{prestori2013gating, cao2022nicotine, ueno2013long, ueno2017coincident}, a process we propose reinforces association strength and promotes generalization across related stimuli. 
In this view, neuromodulator-driven plasticity allows the expansion layer to prioritize salient and behaviorally relevant sensory features, adjusting representations independently of the specific adaptive behaviors implemented downstream.


\section*{Testable predictions for future experiments}\label{sec:predictions}

Experimental evidence and theoretical considerations suggest that plasticity in expansion layers serves complementary roles to associative learning at the output stage. However, to clearly identify these roles, as well as to understand the underlying mechanisms, future experimental work is needed to test specific hypotheses about plasticity in the cerebellum granule layer and MB calyx. In the following, we thus lay out two testable hypotheses that are based on previous observations and theory to clarify the functional role and mechanisms of representation learning in expansion layers.

\subsection*{Prior representation learning improves generalization of adaptive behavior}


A key prediction from representation learning is that plasticity at excitatory input synapses captures statistical relationships between co-occurring stimuli (Box~\ref{info:theory}). 
This raises the question whether unreinforced, input-driven plasticity at excitatory synapses provides a predictive scaffold for later associative learning---a hypothesis that can be addressed by preconditioning paradigms.

In preconditioning, the (repeated) exposure to two stimuli enables both stimuli to become associated with reinforcement, even if only one stimulus is conditioned later on.
This has been observed in flies~\cite{martinez2022higher} and honey bees~\cite{muller2000sensory}, indicating that these species are able to generalize learned associations to statistically related stimuli.
However, the anatomical locus and mechanisms underlying this non-associative learning are not yet understood. 
To test whether expansion layer plasticity enables the learning of stimulus–stimulus associations, preconditioning could be combined with imaging of granule or Kenyon cells as well as their inputs. 
This would allow to verify if expansion layer responses change their tuning to the preconditioned stimuli (and not their inputs), thus localizing the non-associative learning to the expansion layer.

Moreover, the same preconditioning experiment could also address an open question about expansion-layer plasticity. 
Models of predictive learning (Box~\ref{info:theory}) and simulations in the cerebellum~\cite{casali2020cellular} predict that inhibition constrains excitatory plasticity, allowing only a small number of cells to strengthen their synapses while most undergo depression. Thus, after the preconditioning phase, this inhibitory control would produce sharper tuning in a subset of neurons while suppressing activity across the population.


Finally, experiments could selectively knock-out plasticity induction in the expansion layer to causally test its role in preconditioning.
Candidate mechanisms include NMDAR-dependent \calcium{} signaling in granule cells~\cite{pali2024understanding}, and dopamine receptor–mediated pathways in Kenyon cell dendrites~\cite{qiao_input-timing-dependent_2022}. Such perturbations should impair both representational changes and behavioral generalization. These results would identify non-associative input plasticity as a substrate for predictive representation learning—enhancing generalization and improving the efficiency of subsequent associative learning.

\subsection*{Reinforced plasticity enhances the association strength of behaviorally relevant stimuli }
Reinforced plasticity in expansion layers could enhance the association strength of behaviorally relevant stimuli to support adaptive behavior. 
Accordingly, a key prediction is that expansion layer responses are selectively enhanced, and that these plastic changes persist over long timescales and thus contribute to subsequent behavioral adaptation. 

While there is increasing evidence for these predictions in the mushroom body (see Section~\nameref{sec:reinforcedPrinCell}), the cerebellar granular layer remains comparatively understudied in this respect. 
In particular, although it is known that acetylcholine can facilitate potentiation of mossy fiber (MF) synapses onto granule cells~\parencite{prestori2013gating, cao2022nicotine}, it remains open whether associative learning---e.g., mediated by climbing fibers (CF)---also leads to changes in granule cell responses. 
This could be tested with \emph{in vivo} recordings of the granule layer during motor learning or paired MF and CF activation.
An exciting possibility is that reinforcement signals reach the granule layer indirectly, through Golgi-cell disinhibition following CF or Purkinje cell activation~\cite{xu2008climbing, ankri2015novel}, or deep cerebellar nuclei projections to the input stage of the granule layer~\cite{houck2014cerebellar, houck2015cerebellar, gao2016excitatory}.
Strikingly, such nucleocortical connections have been shown to enhance or decrease the amplitude of conditioned eyeblink responses~\cite{gao2016excitatory}.
Demonstrating that these pathways lead to changes in expansion layer responses would clarify whether a layer-specific encoding of association strength and behavioral adaptation represents a conserved computational principle across cerebellum-like circuits.

Moreover, to obtain a better understanding for the behavioral relevance of reinforced plasticity in the expansion layer, more experiments are required that relate behavioral outcomes to plasticity induction in principal cells or interneurons. 
Here, a key prediction is that reinforced plasticity leads to an encoding of association strength in the expansion layer that persists for long timescales, and potentially even independently of subsequent learning at the output layer (see Box~\ref{info:assoc}). 
This could be tested during experiments that compare spontaneous recovery~\cite{yang2023spontaneous} or re-learning~\cite{wang2023forgotten} between wild-type and knock-out mutants with impaired Kenyon cell or granule cell plasticity (e.g., targeting Dop2Rs in \emph{Drosophila} Kenyon cells, or NMDARs in granule cells). 
If the prediction holds, enhanced re-learning or spontaneous recovery rates are expected for wild type due to intact expansion layer plasticity.
Moreover, imaging of expansion layer populations during reversal learning should reveal stable representations of conditioned cues, even as their associated behavior reverses at the output stage. 
Finally, plasticity in the expansion layer should enhance the acquisition of new stimulus contingencies, in line with the involvement of the cerebellum for reward-based reversal learning in humans~\cite{thoma2008cerebellum}.

Testing these predictions would substantially improve our understanding of the role of reinforced expansion layer plasticity in cerebellum-like structures, and reveal potential cellular mechanisms underlying human and animal learning and adaptive behavior.




\section*{Interaction with other processes in cerebellum-like structures}

Although much of this review has focused on plasticity mechanisms operating locally within expansion layers, these mechanisms do not act in isolation. In particular, the following four processes could yield important interactions with expansion layer plasticity to shape adaptive behavior and learning:
(i) developmental and slow homeostatic plasticity;
(ii) heterogeneity and specialization within expansion-layer circuits;  
(iii) recurrent and feedback motifs and 
(iv) plasticity at the input stage. 
Below, we discuss how each of these processes extends the computational role of expansion-layer plasticity.

We focused on plasticity mechanisms in the range of minutes to days that have a direct relation to representation learning. 
However, also slower activity-dependent plasticity and developmental processes are instrumental for the formation of efficient expansion layer representations. 
In the \emph{Drosophila} mushroom body, chronic reduction of projection neuron activity induces compensatory strengthening of PN–KC synapses~\cite{kremer2010structural, pech2015optical}, consistent with homeostatic stabilization of excitatory input, accompanied by parallel homeostatic plasticity in the inhibitory APL neuron~\cite{ apostolopoulou2020mechanisms}.
Moreover, key aspects of the PN-KC connectivity in \emph{Drosophila} form in the absence of odor-receptor induced activity, suggesting that sensory information is not required during the development of the input connections to the expansion layer~\cite{hayashi2022mushroom}.
Therefore, genetically determined developmental processes likely account for most of the structural aspects of the input connections. 
In contrast, representation learning in cerebellum-like circuits likely fine-tunes synaptic transmission strength and neural excitability, although sensory experience can also induce structural remodeling~\cite{kremer2010structural, Baltruschat2020, ruediger2011learning}.
Future modeling studies are needed to address how homeostatic structural plasticity and developmental processes interact with faster plasticity of representation learning.



Moreover, we have worked towards a general perspective on functional principles of plasticity in expansion layers. However, expansion layers are inhomogeneous structures~\cite{cerminara2015redefining}.
In the MB, different types of KCs in the calyx project to different lobes, participate in different memory phases (short, mid- or long-term)~\cite{krashes2007sequential, blum2009short,qin2012gamma} and modulate different types of behavior (e.g., appetitive versus aversive)~\cite{perisse2013different}.
This functional specialization is already evident at the level of the calyx, where $\alpha/\beta$ KCs have been shown to sample predominantly food-related odors, which could facilitate fast learning for ecologically relevant stimuli~\cite{chan2025odour}, whereas 
$\gamma$ KCs are generalists that integrate different odor categories and sensory modalities, have an overall higher number of inputs and a higher firing threshold~\cite{chan2025odour}. 
These properties position $\gamma$ KCs to support flexible representation learning, consistent with observations that spike-timing–dependent plasticity and dopaminergic feedback induce response changes specifically in this population~\cite{qiao_input-timing-dependent_2022, boto2019independent}.
At the same time, other forms of plasticity likely operate in additional KC subtypes, potentially accounting for learning-related changes observed in $\alpha/\beta$ and $\alpha^{\prime}/\beta^{\prime}$ KCs at later memory stages~\cite{yu2006drosophila, akalal2010late}.
A comparable heterogeneity characterizes the cerebellum, which is also organized into microzones that receive different inputs, modulate different deep nuclei and serve distinct functions, thus pointing at a non-uniform structure of the cerebellar cortex~\cite{cerminara2015redefining, gruver2024structured}.
Together, these observations suggest that representation learning is differentially implemented by different types of KCs in the MB or GrCs in different parts of the cerebellar cortex.

In addition, we have only focused on feedforward projections and lateral inhibition in expansion layers. However, these structures also contain recurrent circuit motifs. An important open question is therefore how representation learning in the expansion layer interacts with recurrency, and whether some of the described plasticity mechanisms also apply to these connections. For example, KCs in the calyx form synapses as well as gap junctions to other KCs~\cite{christiansen2011presynapses, liu2016gap}, thus enabling recurrent activity propagation within the expansion layer. 
Interestingly, KCs express type A muscarinic acetylcholine receptors on their dendrites that inhibit odor responses, thus also enabling direct inhibition between KCs~\cite{bielopolski2019inhibitory}. 
This is a setup that is also used in predictive learning theories to enable efficient spike representations in a neural population (Box~\ref{info:theory}). 
Therefore, these synapses might also be subject to balancing LTP/LTD, similar to what has been reported in Golgi cell synapses in the granule layer~\cite{mapelli2016heterosynaptic}.
Moreover, KCs also project to dopaminergic and octopaminergic neurons that in turn target the calyx~\cite{hammer1993identified, sinakevitch2011distribution, lyutova2019reward, boto2019independent}. 
Since these aminergic relevance signals drive further learning in the expansion layer (see Tables~\ref{tab:reinforced_principle_cells} and~\ref{tab:reinforced_inhibition}), this creates a recurrent feedback loop that could enable further increase in association strength, consolidation and generalization of learned associations (Box~\ref{info:assoc}).
A similar recurrent loop exists in the cerebellum, where neurons in the deep nucleus project via mossy fiber inputs to the granule layer~\cite{gao2016excitatory}. 
Since strong MF activation can induce plasticity at MF synapses (Table~\ref{tab:stim_potentiation}), this feedback loop could also be used for enhance associative learning through representation learning in the expansion layer. 
In line with this, the activation of fibers from the nucleus to the granule layer was observed to enhance conditioned eyeblink responses ~\cite{gao2016excitatory}.
Understanding how such recurrent motifs interact with expansion-layer plasticity---and whether these connections are themselves plastic---remains an important open challenge. 

Finally, although the expansion layer transmits input representations to the convergent output layer, upstream input layers also play a critical role in shaping these representations~\cite{pirez2023experience, muscinelli2023optimal}.
In contrast to the expansion layer, the pontine nuclei that project to the cerebellum, as well as PNs that project to the MB calyx form a highly \emph{structured}, \emph{convergent} layer. 
Functionally, this circuit linearly compresses external inputs before they reach the expansion layer~\cite{muscinelli2023optimal} while also supporting other functions such as divisive normalization~\cite{olsen2010divisive}. 
This compression substantially reduces the number of synaptic weights required if external inputs would directly reach the expansion layer~\cite{muscinelli2023optimal}. It is also in line with observation that PN responses are harder to discriminate compared to upstream responses of olfactory sensory neurons~\cite{niewalda2011combined}. 
Yet, despite architectural differences, plasticity mechanisms in PNs and the antennal lobe resemble those observed in the expansion layer: non-associative plasticity leads to a suppression of PNs after prolonged exposure to a stimulus due to inhibitory plasticity~\cite{das2011plasticity, locatelli2013nonassociative}, whereas reinforced plasticity during associative learning enhances AL responses~\cite{faber1999associative}, leading to better discrimination between relevant and irrelevant stimuli~\cite{pirez2023experience} (cf., Figure~\ref{fig:plasticity}D,E).
Although speculative, we argue that the input layer implements essentially the same plasticity as the expansion layer, yielding a division of labor that enhances the efficiency of representation learning. 
This is because implementing plasticity at the input stage requires fewer sites of plasticity and is thus cheaper to implement, whereas representation learning at the expansion layer provides higher specificity.
Representation learning in cerebellum-like circuits may therefore emerge from coordinated plasticity across both input and expansion layers, although how these mechanisms interact across processing stages remains an open question.

\section*{Concluding remarks}

For decades, expansion layers such as the cerebellar granule layer and the insect mushroom body calyx have been viewed as static substrates for pattern separation—structures that orthogonalize sensory inputs to minimize interference during downstream associative learning \cite{marr1969theory,albus1971theory}. 
Accumulating evidence shows that these layers are plastic, integrating non-associative and reinforcement-driven mechanisms (Tables~\ref{tab:stim_potentiation}-\ref{tab:reinforced_inhibition}) to refine sensory representations before associative learning occurs. 
This shift reframes expansion layers as \textit{active sites of representational learning} where saliency and behavioral relevance are encoded upstream of output layers, thus challenging classical assumptions that restrict learning to downstream synapses (see Box~\ref{info:assoc}).

Moving forward, it will be key to determine how diverse plasticity processes interact during learning to balance discrimination, generalization and behavioral relevance.
Here, a close interaction between experiments and theoretical approaches that integrate expansion-layer plasticity into normative frameworks of predictive coding and reinforcement learning (see Box~\ref{info:theory}) could yield a better functional understanding, guide future experiments and extend insights across diverse neural systems.
Finally, understanding the principles of representation learning in cerebellum-like structures could bridge biological and artificial intelligence, revealing how neural networks achieve rapid learning for adaptive behavior. 

\section*{Acknowledgments}
We would like to thank Gaia Tavosanis, André Fiala, Samuel Muscinelli, and Bertram Gerber for inspiring discussions on plasticity in the MB calyx and its behavioral relevance. We would also like to thank Bertram Gerber and Irene Pellini for helpful feedback on the manuscript.
L.R., F.M. and V.P. received support by the Max Planck Society. 
L.R. was funded by the German Research Foundation (DFG) as part of the SPP 2205 (project number 430157073) and via the SFB 1286 - “Quantitative Synaptology”.
V.P. received support from the SFB 1528 - “Cognition of Interaction”.
A.F.C. was funded by a Daimler Benz foundation grant. A.F.C. also thanks support from core funding from the Chair of Computational Neuroscience at the School of life Sciences at the Technical University of Munich.
The funders had no role in study design, data collection and analysis, decision to publish, or preparation of the manuscript.





\printbibliography

\end{document}